\documentclass[preprint,showpacs,preprintnumbers,amsmath,amssymb]{revtex4}
\usepackage[dvips]{graphicx}
\usepackage{natbib}
\usepackage{bm}

\def\hf{{\frac{1}{2}}}
\begin{document}

\bibliographystyle{plainnat}
\title{Statistical Equilibrium of trapped slender vortex filaments - a continuum model}
\author{Timothy Andersen}
\email{andert@rpi.edu}
\author{Chjan Lim}
\email{limc@rpi.edu}
\affiliation{Mathematical Sciences, RPI, Troy, NY, 12180}

\date{\today}

\begin{abstract}
Systems of nearly parallel, slender vortex filaments in which angular
momentum is conserved are an important simplification of the Navier-Stokes
equations where turbulence can be studied in statistical equilibrium.  We
study the canonical
Gibbs distribution based on the Klein-Majda-Damodaran (KMD) (\cite{Klein:1995}) model and 
find a divergence in the mean square vortex position from that of the point vortex model of
 \cite{Onsager:1949} at moderate to high temperature.  We subsequently develop a free-energy
equation based on the non-interacting case, with a spherical constraint, which we approximate using
the method of Kac-Berlin \cite{Berlin:1952}, adding a mean-field term for
logarithmic interaction.  We use this free-energy equation to predict the Monte Carlo results.
\end{abstract}

\maketitle

\section{Introduction}
In statistical equilibrium the behavior of collections of nearly parallel slender vortex
filaments, periodic in the z-direction and confined by angular momentum in
the xy-directions,  is understood to be nearly 
identical to that of point vortices (in which no z-direction exists) of Onsager \cite{Onsager:1949} for a
wide range of temperatures \footnote[1]{By temperature
we refer to the Lagrange multiplier for the energy of the vortex system and
not the molecular temperature that a thermometer measures.}.  
When a filament is quite straight, its internal configuration has a much smaller influence on
macroscopic statistical properties of the system as a whole (e.g. density and shape of a bundle of filaments) than the mean position of its center-line
relative to the center-lines of the other filaments in the bundle.  Taking
advantage of this fact, \cite{Lions:2000} were able to derive 
a mean-field theory for the density of a system of
$N$ nearly parallel vortex filaments using the asymptotically derived
PDE of Klein-Majda-Damodaran (KMD) \cite{Klein:1995},\cite{Ting:1991}, adding mathematical rigor
to the statistical equilibrium models for turbulence that Chorin has pioneered \cite{Chorin:1994}.
However, no adequate theory yet exists for the behavior of the KMD system
at moderate-to-high temperature where turbulent behavior is more extreme and
long-range order is minimal.  Therefore, 
we must rely on computational methods to explore this important regime.  
Using the Path Integral Monte Carlo method of \cite{Ceperley:1995},
originally created to study quantum bosons, we
sample the Gibbs canonical ensemble for the Hamiltonian that \cite{Klein:1995}
have derived.  
Our findings show that the point vortex
analogy indeed breaks down at high temperatures.  While the density of point 
vortex systems increases monotonically with increasing temperature, we show that
the density of the three-dimensional system actually \emph{decreases} at a particular temperature.

As Nordborg and Blatter point out \cite{Nordborg:1998}, there is a close
relationship between the quantum mechanics of bosons in (2+1)-D and the behavior of vortex filaments.  Using methods developed for quantum field theory to evaluate functional integrals, we derive a mean-field free-energy valid for
all temperatures that predicts the behavior of filaments represented by smooth curves (a continuum model).

\section{Model}
\subsection{Slender Vortex Filaments}
A system of $N$ Nonlinear Schroedinger Equations (NLSEs) describes the time-evolution of slender vortex filaments
\begin{equation}
-i\partial_t\psi_j = \alpha\partial_{\sigma\sigma}\psi_j + 
\frac{1}{2}\sum_{j\neq k}{\frac{\psi_j - \psi_k}{|\psi_j - \psi_k|^2}},
\end{equation} where $\psi_j(\sigma,t) = x_j(\sigma,t) + iy_j(\sigma,t)$ is the position of vortex $j$ at
position $\sigma$ along its length at time $t$, and $\alpha$ is the core structure constant
 (\cite{Klein:1995}).  Vortex strengths are assumed to all be the same and are set to unity.  The
position in the complex plane,
$\psi_j(\sigma,t)$, is assumed to be periodic in $\sigma$ with period $L$.

This system of PDEs can be expressed as a Hamiltonian system
\begin{align}
H_N = \alpha\int_0^L\sum_{k=1}^{N} \frac{1}{2}\left|\frac{\partial \psi_k(\sigma)}{\partial\sigma}
\right|^2 d\sigma\nonumber - \int_0^L\sum_{k=1}^N\sum_{i>k}^N \log|\psi_i(\sigma) - \psi_k(\sigma)|d\sigma.
\end{align}
  To this we also add a trapping potential which
conserves angular momentum,
\begin{equation}
I_N = \int_0^L\sum_{k=1}^N |\psi_k(\sigma)|^2 d\sigma.
\end{equation}

For our simulations we assume that the filaments are piecewise linear, divided into
an equal number of segments of equal length.  This discretization leads to the Hamiltonian,

\begin{align}
H_N(M) = \alpha\sum_{j=1}^{M}\sum_{k=1}^{N} \frac{1}{2}\frac{|\psi_k(j+1) -
\psi_k(j)|^2}{\delta}\nonumber - \sum_{j=1}^{M}\sum_{k=1}^N\sum_{i>k}^N
\delta\log|\psi_i(j) - \psi_k(j)| ,
\end{align}

and angular momentum
\begin{equation}
I_N = \sum_{j=1}^{M}\sum_{k=1}^N \delta|\psi_k(j)|^2
,\end{equation} where $\delta$ is the length of each segment and $M$ is the number of segments.  
For purposes of later discussion, the point where two segments meet is called a ``bead"
in PIMC terminology.

\subsection{Path Integrals and Partition Functions}
The path integral method of Feynman in its imaginary time density matrix format involves
evaluating the Gibbs measure of a set of paths,
\begin{equation}
G_N(M) = \frac{\exp{\left(-\beta H_N(M) - \mu I_N(M)\right)}}{Z_N(M)}
,\end{equation} where
\begin{equation}
Z_N(M) = \sum_{all paths} G_N(M).
\end{equation}  The Gibbs probability,
used in the \cite{Klein:1995} model by \cite{Lions:2000}, gives a probability for a path, and that allows us to use the path integral method.  Originally
developed for quantum systems of bosons in imaginary time, PIMC applies to the vortex model perfectly as shown by \cite{Nordborg:1998}.

Although we use most of Ceperley's original PIMC methods, there is at least one major difference in notation.  
In quantum path integral computations $\beta$ becomes the imaginary time length of the path.  Since
we are modeling real filaments with their own periodic length, $L$, it is important
to understand that our $\beta$ corresponds to $1/\hbar$ and not time.

\section{Numerical Results}
This section describes our numerical results based on the model and method described above.
Filaments were initialized by scattering them with uniform randomness in a square
of side $10$.
Monte Carlo moves involved choosing first a vortex filament to change with uniform
randomness, then choosing the type of move, either moving the entire chain or rearranging
the internal configuration via bisection.  If moving the entire chain, a new point
was chosen within a square with a side-length of 10, centered on at the current filament's
xy-planar position.  

Energy was calculated the same way for both types of moves, using
the multilevel method of \cite{Ceperley:1995}.  Therefore, even the wholechain move had
the possibility of being rejected before energy was fully evaluated.

\begin{figure}[tph]
\begin{center}
\includegraphics[width = \textwidth]{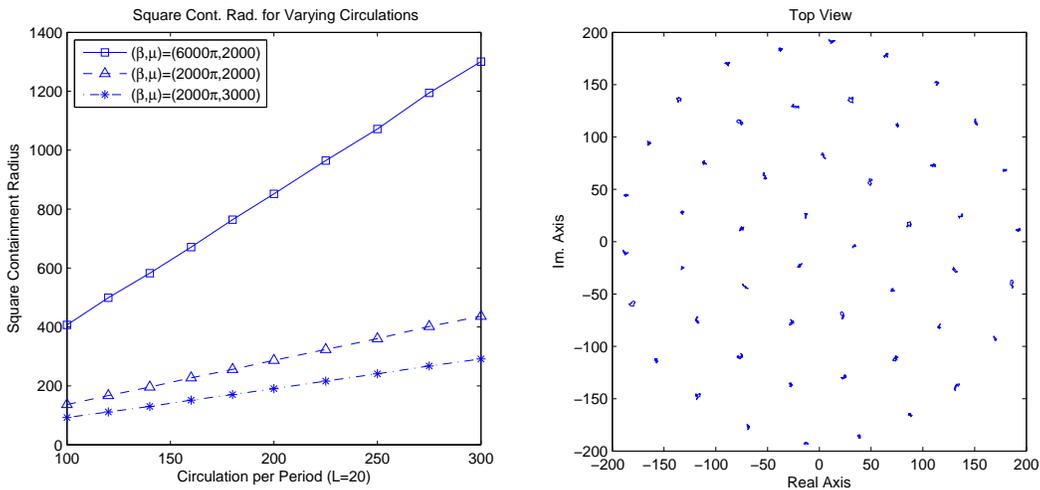}
\end{center}
\caption{Low-temperture ensures a well-defined containment radius.}
\label{fig:fig1}
\end{figure}

We ran our Monte Carlo simulations until energy settled down to a steady mean.  For the high $\beta$ (i.e. low temperature) simulations, we see triangular lattices
form upon convergence.  Although the fluid remains in a liquid state, these lattices
have a crystaline structure resembling that of a solid in which the filaments
vibrate but maintain a fixed position w.r.t. their neighbors (Figure \ref{fig:fig1}, right).

Our first finding is a confirmation of findings in \cite{Lim:2005} which demonstrate
the relationship between square containment radius, $R = \langle \max_{i}(|\psi_i|^2)\rangle$, 
and the parameters $\beta$
and $\mu$ and the circulation $\Omega$, where $\alpha$ was allowed to remain fixed.
We find near perfect agreement in the line slopes to \cite{Lim:2005} formula for
square containment radius $R^2=N\beta/(4\pi\mu)$.  Additionally, the Hamiltonian used in
\cite{Lim:2005} is divided by $\pi$, which is not done in \cite{Klein:1995}.  
Therefore, the formula for our square
containment radius is 
\begin{equation}
R^2=N\beta/(4\mu),
\end{equation}
which gives the slope equal to $\beta/(4\mu)$ in agreement
Figure \ref{fig:fig1} left.

\begin{figure}[bpth]
\begin{center}
\includegraphics[width = \textwidth]{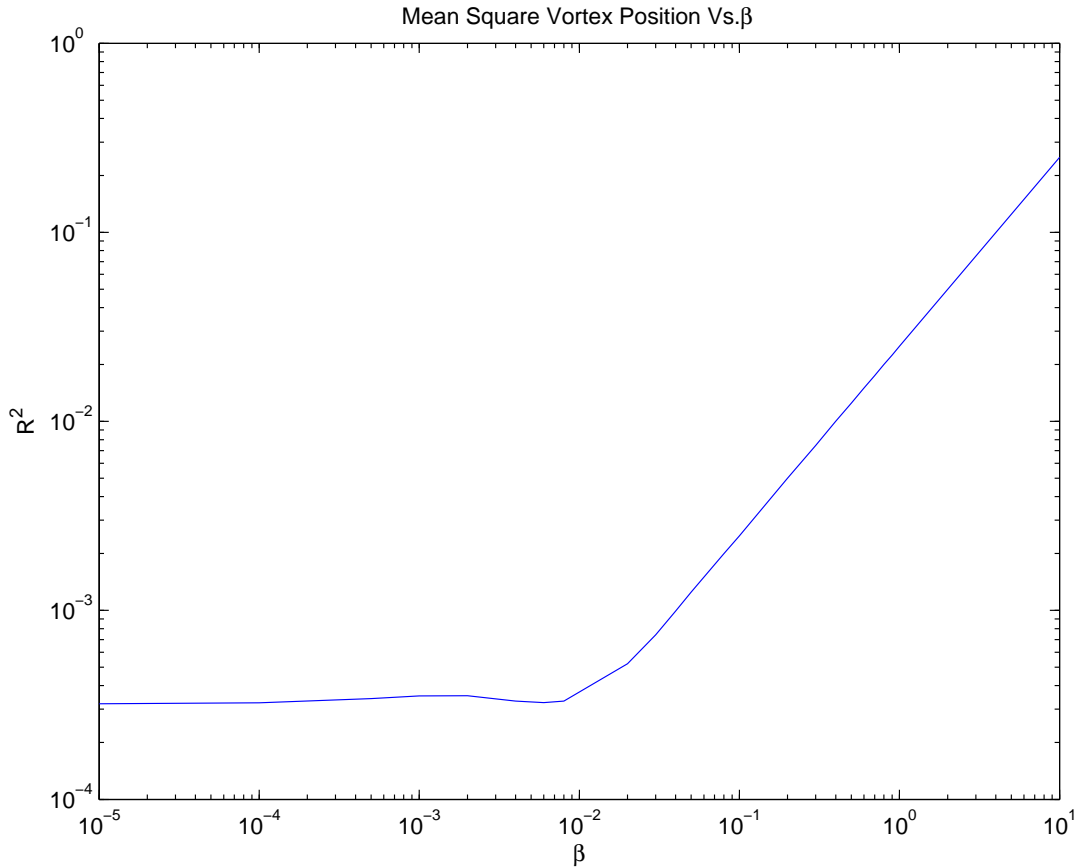}
\end{center}
\caption{At a specific temperature ($\beta\sim0.004$) the entropy takes over from logarithmic interaction as the main force for expansion.  Here $L=100$, $\alpha=10^6$, $\mu=2000$,
$N=200$, and $M=64$.}
\label{fig:fig2a}
\end{figure}

\begin{figure}
\begin{center}
\includegraphics[width = \textwidth]{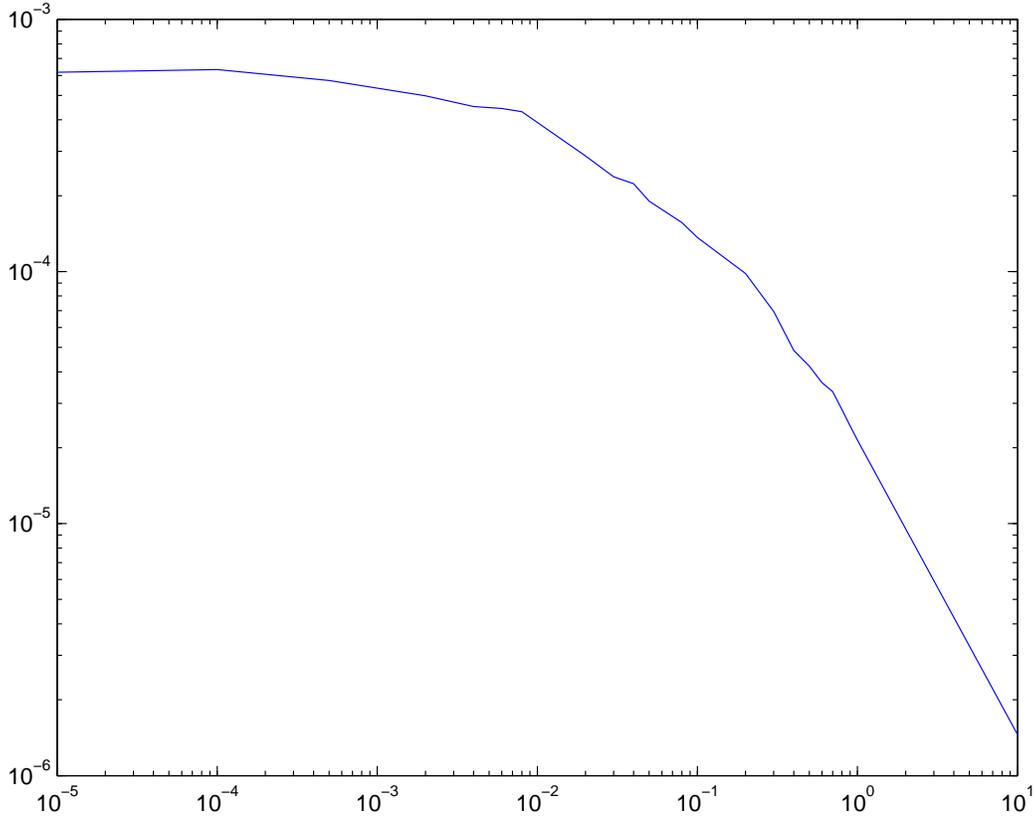}
\end{center}
\caption{The amplitude is much less than the length $L=100$.}
\label{fig:fig2b}
\end{figure}

Our next finding, shown in Figure \ref{fig:fig2a}, is much more important as 
it deviates from the formula of \cite{Lim:2005}.  Here we see that the pure interaction of the
trapping potential term and the logarithmic interaction term is no longer valid.  While the asymptotic
assumptions of the model are not broken,
\begin{equation}
L \gg A^2 \gg \epsilon,
\end{equation} where $A$ (Figure \ref{fig:fig2b}) is the amplitude of the
vortices and $\epsilon$ is the core size (here assumed to be small), the mean-field behavior
at high $\beta$ breaks down.  At the point where the slope of the curve changes around $\beta\sim 0.004$, the entropy begins to affect the statistical equilibrium $R^2$ of the bundle, reducing the
slope.  As we will
see in Section 4, the discretization has a profound lowering effect on this part of the curve, making a
continuum model necessary for accurate prediction of the entropy regime $\beta<0.004$.


\section{Free Energy Theory}
While the above results are interesting of themselves, our investigation is incomplete without a 
free-energy theory to explain them for all choices of parameters.  Directly calculating
the free-energy of the interacting system is impossible with current mathematical knowledge.  However,
we might approximate the behavior of $R^2$ under a mean-field interaction term if we place the filaments in
the non-interacting system on average at an undetermined distance $R$ from the center via a spherical constraint.  Quantum mechanical path integral methods are particularily useful for this section.

The classical mean-field action (in imaginary time causing the potential to flip sign) is
\begin{equation}
 S_{mf} = \int_0^L d\tau -\hf \alpha\beta|\partial_\tau \psi(\tau)|^2 - \mu|\psi(\tau)|^2 + \beta N\log(R^2)/4,
\end{equation} and the spherical constraint, we enforce with a delta function
\begin{equation}
 \delta[|\psi(\tau)|^2 - R^2],
\end{equation} which has integral representation
\begin{equation}
 \int_{-\infty}^{\infty}\frac{d\sigma}{2\pi} \exp -i\sigma[|\psi(\tau)|^2 - R^2],
\end{equation} at each $\tau$.

Now let
\begin{equation}
 S_{mf}^{sphere} = \int_0^L d\tau -\hf \left[\alpha\beta|\partial_\tau \psi(\tau)|^2 + (i\sigma + 2\mu)|\psi(\tau)|^2 - 2\beta N\log(R^2)/4 - iR^2\sigma(\tau)\right],
\end{equation} be the action.

The non-dimensional free-energy ($f = \beta F$, where $F$ is in energy units) as a function of $i\sigma(\tau)$ is then
\begin{equation}
 f[i\sigma] = -\ln\int D\psi \exp S_{mf}^{sphere},
\end{equation} where $D\psi$ represents the functional integral over all paths $\psi(\tau)$ and the
partition function for $N$ filaments is
\begin{equation}
 Z_N = \int D\sigma \exp(-Nf[i\sigma])
\end{equation} \cite{Hartman:1995}.  (The exchange of the functional integration of $\sigma$ with that of $\psi$ is permitted in this case because the action is negative definite.) Clearly as $N\rightarrow\infty$ the saddle point gives the main contribution
\cite{Berlin:1952}.  Therefore, 

\begin{equation}
 f_\infty = \lim_{N\rightarrow\infty}-\frac{1}{N}\ln Z_N = f[\eta],
\end{equation} where $\eta = i\sigma(\tau)$ is the saddle point.  We let $\beta' = \beta N$ and $\alpha' = \alpha N^{-1}$ to take the non-extensive limit while keeping $\alpha\beta = \alpha'\beta'$ constant.

The free-energy involves a simple harmonic oscillator with a constant external force

\begin{equation}
 f[i\sigma] = - \ln h[i\sigma] + \int_0^L d\tau -\hf i\sigma R^2 - \beta' \log(R^2)/4.
\end{equation}  
Here $h$ is the partition function for a quantum harmonic oscillator in imaginary time,

\begin{equation}
 h[i\sigma] = \int D\psi \exp \int_0^L d\tau -\hf m[|\partial_\tau \psi|^2 + \omega^2|\psi|^2] ,
\end{equation} which has the well known solution for periodic paths in (2+1)-D where
we have integrated end-points over the whole plane,
\begin{equation}
 h[i\sigma] = \frac{ e^{-\omega L} } { \left(e^{-\omega L} - 1\right)^2 },
\end{equation} where $m=\alpha'\beta'$ and $\omega^2 = (i\sigma + 2\mu)/(\alpha'\beta')$ \cite{Brown:1992}, \cite{Zee:2003}.

Let us first make a change of variables $\lambda = i\sigma + 2\mu$.  Then the free-energy reads
\begin{equation}
 f[\lambda] = \left[\int_0^L d\tau (\mu - \hf\lambda)R^2\right]  - \beta' L\log(R^2)/4 - \ln\frac{e^{-\omega L }}{\left(e^{-\omega L} - 1\right)^2},
\end{equation} 
where $\omega = \sqrt{\lambda/(\alpha'\beta')}$.  

Letting $\eta$ be the saddle point and
independent of $\tau$ \cite{Hartman:1995}, we find that the free-energy is
\begin{equation}
 f[\eta] = (\mu - \hf\eta)LR^2 - \beta' L\log(R^2)/4 + \sqrt{\frac{\eta}{\alpha'\beta'}} + 2\log\left|\exp\left({-\sqrt{\frac{\eta}{\alpha'\beta'}}L}\right) - 1\right|,
\end{equation} where $\partial f/\partial \lambda = 0$ and $\partial f/\partial R^2 = 0$ at $\eta$ and $R^2$.

Since
\begin{equation}
 \frac{\partial f}{\partial R^2} = (\mu - \lambda/2)L - L\beta'/(4R^2),
\end{equation}
\begin{equation}
 R^2 = \frac{\beta'}{4(\mu - \lambda/2)}.
\label{eqn:rsq}
\end{equation}

Substituting for $R^2$ we get
\begin{equation}
 f[\lambda] = \beta' L/4 + \sqrt{\frac{\lambda}{\alpha'\beta'}}L + 2\log\left|\exp\left({-\sqrt{\frac{\lambda}{\alpha'\beta'}}L}\right) - 1\right| - \frac{\beta' L}{4}\log\frac{\beta'}{4(\mu - \lambda/2)}.
\end{equation}

Taking the derivative
\begin{equation}
 \frac{\partial f}{\partial \lambda} = \frac{L}{2\sqrt{\alpha'\beta' \lambda}} - \frac{L\exp\left({-\sqrt{\frac{\lambda}{\alpha'\beta'}}L}\right)}{\sqrt{\alpha'\beta' \lambda}\left|\exp\left({-\sqrt{\frac{\lambda}{\alpha'\beta'}}L}\right) - 1\right|} - \frac{L\beta'}{8(\mu - \lambda/2)},
\label{eqn:dfdlambda}
\end{equation} we get a transcendental equation that cannot be solved analytically.  However, it
is clear that $\lambda=2\mu$ and $\lambda=0$ are transition points.  For the rest of the paper we will
drop the primes from $\beta'$ and $\alpha'$.

This system is a quantum one in all but name with an inverse quantum temperature of $L$.  To study any phase transitions, we must take the $L\rightarrow \infty$ limit because quantum phase transitions \emph{only} occur at absolute zero \cite{Sachdev:1999}.  Therefore, taking the limit we get a per unit length energy of
\begin{equation}
 f_{grnd}[\eta] = \frac{\beta}{4} + \sqrt{\eta/(\alpha\beta)} - \frac{\beta}{4}\log\left(\frac{\beta}{4(\mu - \eta/2)}\right),
\end{equation} where
\begin{equation}
 \eta = 2\mu - \frac{1}{8}\beta(-\beta^2\alpha \pm \sqrt{\beta^4\alpha^2 + 32\alpha\beta\mu}),
\end{equation} of which we take
\begin{equation}
\eta = 2\mu - \frac{1}{8}\beta(-\beta^2\alpha + \sqrt{\beta^4\alpha^2 + 32\alpha\beta\mu})
\end{equation} as giving physical results (shown in Figure \ref{fig:fig2}.)

The two points where the free-energy is non-analytic,
\begin{equation}
 \eta = 0,\quad 2\mu,
\end{equation}  correspond to
\begin{equation}
 \beta = - \sqrt[3]{\frac{32\mu}{\alpha}},\quad 0,
\label{eqn:betatrans}
\end{equation} which we will call $\beta_{-c}$ and $\beta_{c}$ respectively.

The change in the behavior of $R^2$ we observe in our Monte Carlo results in Figure \ref{fig:fig2} at $
\beta\sim 10^{-2}$ is not a phase transition but a smooth transition, since the free-energy is
smooth at all $\beta$ other than $\beta_{-c}$ and $\beta_{c}$.  

\begin{figure}[tph]
\begin{center}
\includegraphics[width = 0.5\textwidth]{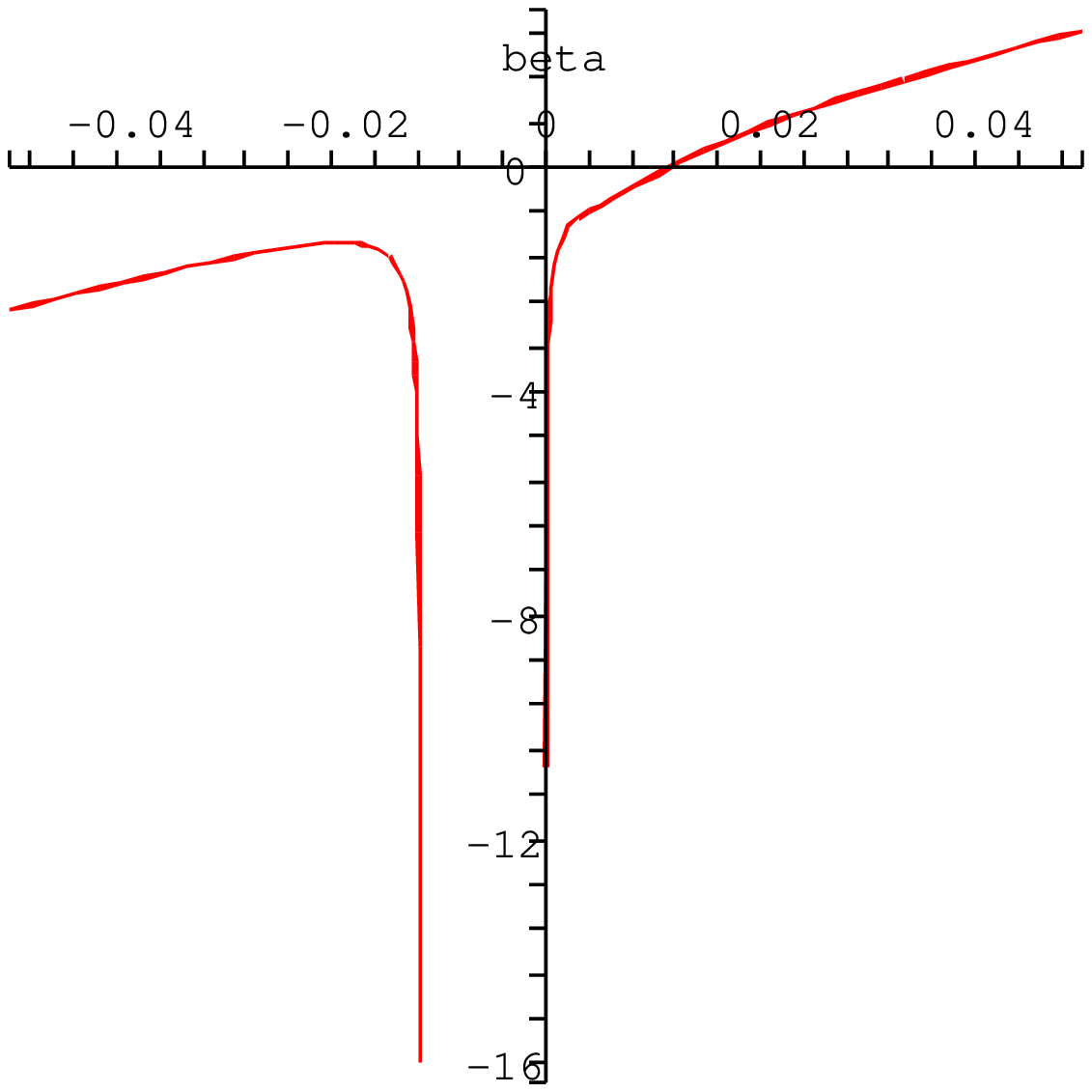}
\end{center}
\caption{The specific heat in Equation \ref{eqn:spheat}, shown here with parameters $\alpha=10^6$,
$N=200$, and $\mu=2000$ diverges at $\beta=0$ and $\beta=- \sqrt[3]{\frac{32\mu}{\alpha}}$.
For $\beta$ in this interval it has complex value.}
\label{fig:spheat}
\end{figure}

\begin{figure}[tph]
\begin{center}
\includegraphics[width = \textwidth]{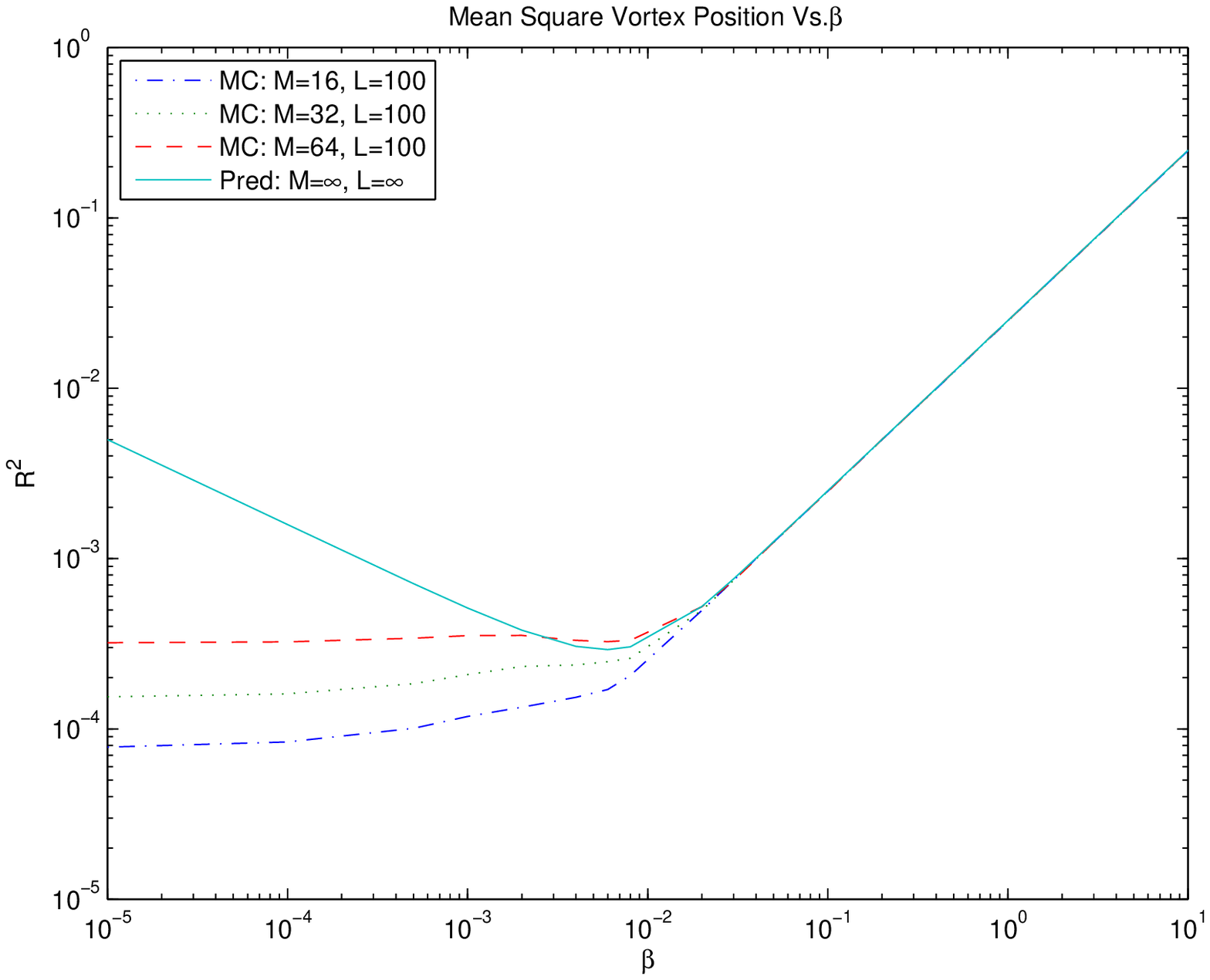}
\end{center}
\caption{As the broken segment model used in the Monte Carlo simulations approaches the contiuum limit $M=\infty$, the change in $R^2$ with $\beta$ is predicted to approach a ``v''-shaped (corresponding to
a polynomial function in $\beta$) curve.}
\label{fig:fig2}
\end{figure}

The specific heat
\begin{equation}
 C = -\beta^2\frac{\partial^2 f_{grnd}}{\partial \beta^2}
\label{eqn:spheat}
\end{equation} \cite{Berlin:1952} (see Figure \ref{fig:spheat}) can be represented in series as
\begin{equation}
 C = -\frac{3}{4}\sqrt{2\mu/\alpha}\beta^{-1/2} + O(higher),
\end{equation} indicating a critical exponent of $\nu=-1/2$, suggesting a second-order (continuous) phase transition.  \footnote{Despite the quantum analogy, this is not the lambda transition of superfluids.  These particles are Boltzmannons not Bosons (no permutations of particles) and no percolation can occur.  If the logarithmic term were taken away, there would be no phase transition at $\beta=0$.}

\section{Conclusion}
To our knowledge no one has done Monte Carlo simulations of the system proposed 
in \cite{Lions:2000}.  Excellent simulations have been done in cases 
where boundaries are periodic in all directions such as \cite{Nordborg:1998} and \cite{Sen:2001} using 
the London energy functional for flux line lattices, which differs from that of \cite{Klein:1995} only 
in that the interaction potential is a modifed Bessel's function (log-like at short distances).
  However, 
 free boundary conditions with the addition of the conservation of angular
momentum make this problem different and specifically applicable to
fluid statistics and to a continuous filament model of the Gross-Petaevskii equation in the Thomas limit --
details of which will appear in a later paper.

Our findings -- comparing the point vortex expression for $R^2$ with \ref{eqn:rsq} -- indicate a special regime of temperature where entropy of the
filaments has a profound effect on variance of vortex position, which has not been observed before.  Also new is the discovery of a quantum, i.e. infinite  $L$, phase transition in the continuous vortex filaments model at zero value of inverse temperature as well as a negative $\beta$ transition in Equation \ref{eqn:betatrans}.  According to Sachdev \cite{Sachdev:1999} characteristics of the quantum transition extend to finite values of $L$.  Hence, our comparison between theoretical values for $R^2$ at infinite $L$ and PIMC values of $R^2$ at finite $L$ are valid.

We developed a theory based on a combination of the non-interacting case with a mean-field
estimate for logarithmic interaction.  The resulting free-energy allows us to predict 
the $\beta$ point where entropy forces $R^2$ to increase with temperature and becomes the dominant resistance to the angular momentum rather than interaction.  We have also shown that
this is not a phase transition for this model.  More interesting behavior may be observed at negative $\beta$.  We also plan to apply our methods to periodic boundaries.

Lions and Majda \cite{Lions:2000} have already suggested applicability of their
novel derivations in the area of geophysical and astrophysical convection 
such as \cite{Julien:1996} have modeled.  We say that our results are equally
applicable and we may extend them to Bose-Einstein Condensates in the future.

\centering
\large
{\bf Acknowledgments}
\flushleft
\normalsize
This work is supported by ARO grant W911NF-05-1-0001 and DOE grant 
DE-FG02-04ER25616. We acknowledge the scientific support of Dr. Chris Arney, Dr. Anil Deane and Dr. Gary Johnson.
\pagebreak
\bibliography{pimcposter}
\end{document}